\documentclass[aps,prb,preprint,superscriptaddress,longbibliography]{revtex4-2}

\usepackage{amsmath,amssymb,bm}
\usepackage{graphicx}
\usepackage{subcaption} 
\usepackage{hyperref}
\usepackage{xcolor}
\usepackage{booktabs}

\hypersetup{hidelinks}

\begin{document}

\title{Paraexciton Excitation in Cu$_2$O under Laguerre--Gaussian Illumination}

\author{Nguyen Que Huong}
\email{nguyenh@marshall.edu}
\affiliation{Department of Mathematics and Physics, Marshall University, Huntington, West Virginia 25755, USA}
\author{David W. Facemyer}
\email{facemyer3@marshall.edu}
\affiliation{Department of Mathematics and Physics, Marshall University, Huntington, West Virginia 25755, USA}


\date{\today}

\begin{abstract}
In Cu$_2$O the lowest yellow exciton, the $\Gamma_2^+$ paraexciton, is optically inaccessible in conventional spectroscopy because transitions to this state are forbidden in both electric-dipole and electric-quadrupole approximations. We investigate whether optical fields carrying orbital angular momentum (OAM) can overcome this restriction. A microscopic symmetry analysis identifies the gradient-assisted $l=5$ and direct $l=6$ OAM channels as the leading contributions that couple to the paraexciton, independent of the detailed radial profile of the optical field. Calculations for finite-waist Laguerre--Gaussian beams, however, show that the corresponding matrix elements are strongly suppressed because the optical field varies only weakly over the exciton Bohr radius. Thus, satisfying the OAM selection rule alone is insufficient: efficient excitation requires not only the correct angular symmetry but also optical-field variations on the spatial scale of the exciton. This second condition is achieved by localized OAM fields. Expressing the coupling in terms of the physical intensity-ring radius provides a direct comparison between the optical and excitonic length scales and reveals that the optimal localization is determined primarily by the polynomial degree of the target cubic harmonic. For the degree-six $\Gamma_2^+$ paraexciton the strongest coupling occurs for an intensity-ring radius of approximately $6a_B$--$7a_B$. These results establish that paraexciton excitation is governed jointly by symmetry and spatial localization: cubic symmetry selects the allowed OAM channels, whereas the polynomial degree sets the characteristic radial scale for efficient coupling. This work provides both the symmetry framework and a practical design rule for engineering structured-light excitation of paraexcitons in Cu$_2$O.

\end{abstract}

\keywords{Cu$_2$O, paraexciton, Laguerre--Gaussian beam, orbital angular momentum, structured light}

\maketitle

\section{Introduction}

Although clear evidence for excitons in Cu$_2$O appeared in the optical absorption spectrum nearly 80 years ago, with the first observations reported in the early work of Hayashi and Katsuki~\cite{HAY1}, subsequent optical studies established the hydrogen-like structure of the yellow and green absorption series~\cite{hayashi1959}. These excitons continue to attract considerable attention because of the unusual band structure of the crystal. The rich exciton spectrum of Cu$_2$O has also enabled detailed studies of highly excited states and interseries optical transitions, with potential applications to coherent exciton manipulation and nonlinear optics \cite{rommel2021interseries}.

The cubic lattice of Cu$_2$O belongs to the $O_h$ point group. In the valence band, the five $3d$ orbitals are split by the crystal field into a threefold-degenerate $\Gamma_5^+$ level and a lower twofold $\Gamma_3^+$ level. When spin is included, the highest valence level becomes a doubly degenerate $\Gamma_7^+$ band and a fourfold-degenerate $\Gamma_8^+$ band, the latter lying approximately $130\,\mathrm{meV}$ lower in energy~\cite{KAV1}. In the conduction band, the $4s$ orbital forms the lowest conduction band with $\Gamma_1^+$ orbital symmetry, which becomes the doubly degenerate $\Gamma_6^+$ representation when spin is included. A recent comprehensive review summarizes the experimental and theoretical understanding of the yellow-exciton spectrum and its fine structure in Cu$_2$O \cite{heckotter2025}.

The $1S$ yellow exciton is formed from an electron in the lowest conduction band, $\Gamma_6^+$, and a hole in the highest valence band, $\Gamma_7^+$, across an energy gap of approximately $2.173\,\mathrm{eV}$. Electron--hole exchange splits the fourfold yellow-exciton manifold into a higher-lying, triply degenerate $\Gamma_5^+$ orthoexciton and a lower-lying, nondegenerate $\Gamma_2^+$ paraexciton. The paraexciton is the lowest excitonic state and lies approximately $12\,\mathrm{meV}$ below the orthoexciton.

Because the conduction and valence bands involved in the yellow exciton have the same parity, electric-dipole excitation is forbidden. The effects of uniaxial stress on excitons in Cu$_2$O have therefore played an important role in studies of the exciton structure. Waters \textit{et al.} used Raman scattering to investigate the yellow exciton under uniaxial stress and observed both orthoexciton splitting and a Raman shift~\cite{WAT1}. The stress-induced energy shift of the paraexciton was measured by luminescence~\cite{MYS1}, and the exciton energy levels and their splittings were also treated theoretically~\cite{TRE1}.

The $\Gamma_5^+$ orthoexciton is allowed through electric-quadrupole absorption and was first observed in 1960~\cite{GROS1}. By contrast, the $\Gamma_2^+$ paraexciton remains forbidden in both electric-dipole and electric-quadrupole transitions. The optical inactivity of the paraexciton also gives the dark states comparatively long lifetimes, making them of interest for exciton accumulation and collective phenomena \cite{farenbruch2020l}.

The paraexciton can nevertheless become observable when a magnetic field mixes it with the quadrupole-allowed orthoexciton \cite{farenbruch2020b}. The same work discusses two ways of making the nominally forbidden transition observable: magnetic-field-induced mixing of the paraexciton with the $M=0$ orthoexciton and a two-photon excitation mechanism without an external perturbation involving an electric-quadrupole--magnetic-dipole process.

Because direct paraexciton luminescence is strongly suppressed, a BEC of $1S$ paraexcitons was instead detected by mid-infrared induced-absorption imaging in bulk Cu$_2$O below 400 mK~\cite{Morita2022}, using the internal $1S$--$2P$ transition.

We notice that direct optical access to the $\Gamma_2^+$ paraexciton state requires a transition carrying higher orbital angular momentum than is available in an ordinary plane-wave-like optical field.

Laguerre--Gaussian (LG) beams are well-known solutions of the paraxial wave equation. Since their early development~\cite{KOG1}, they have become standard tools in modern optics and have received renewed attention through applications such as optical trapping~\cite{jesacher2004}. As higher-order beams, LG modes can transfer orbital angular momentum (OAM) to matter and generate optical vortices. A particularly useful feature is that LG modes can, in principle, be prepared with arbitrarily large integer winding number.

In this work, an LG beam is applied to Cu$_2$O and its effect on the yellow-exciton manifold, especially the paraexciton, is considered. The central symmetry result is that transitions to the paraexciton become allowed for particular orders of the LG beam. The original derivation of these channels is developed first. We then extend the analysis numerically to finite-waist beams and localized OAM fields in order to determine whether the symmetry-allowed transitions remain appreciable once the spatial scale of the exciton is taken into account.

The paper is organized as follows. Section~\ref{sec:interaction} introduces the light--matter interaction Hamiltonian. Section~\ref{sec:paraexciton} constructs the paraexciton state and identifies the $l=5$ and $l=6$ LG channels. Section~\ref{sec:matrix_element} gives the transition matrix element in the exciton basis. Sections~\ref{sec:finite_lg}--\ref{sec:degree_scaling} develop the finite-waist and localized-OAM analysis, and Sec.~\ref{sec:discussion} discusses the physical interpretation and implications.

\section{Interaction Hamiltonian for a Laguerre--Gaussian Beam}
\label{sec:interaction}

We consider a Cu$_2$O crystal illuminated by a Laguerre--Gaussian beam. The lowest conduction band has $\Gamma_6^+$ symmetry and the highest valence band has $\Gamma_7^+$ symmetry. It is convenient to use the second-quantized electron--hole formalism, with the electron and hole states expanded in Bloch functions of the periodic lattice.

The interaction Hamiltonian between the electromagnetic field and an electron is written in the usual minimal-coupling form,
\begin{equation}
H_{\mathrm{int}}
=
-\frac{e}{2m_0c}
\left(
\mathbf A\cdot\mathbf p
+
\mathbf p\cdot\mathbf A
\right)
+
\frac{e^2}{2m_0c^2}\mathbf A^2,
\label{eq:minimal_coupling}
\end{equation}
where $e$, $m_0$, and $\mathbf p$ are the electron charge, free-electron mass, and momentum operator, respectively. The positive-frequency part of the vector potential of an LG beam is written as~\cite{LOU1}
\begin{equation}
\mathbf A_{l}^{(+)}
=
A_0
\left(
\alpha\mathbf e_x+\beta\mathbf e_y
\right)
U_{kl}(\mathbf r)
\exp(-i\omega t+ikz)
\hat C_{\mathbf K,\alpha,\beta},
\label{eq:vector_potential}
\end{equation}
where $A_0$ is a complex amplitude, $\alpha\mathbf e_x+\beta\mathbf e_y$ is the polarization vector, $|\alpha|^2+|\beta|^2=1$, and
\begin{equation}
k=\eta(\omega)\frac{\omega}{c}.
\end{equation}
The operator $\hat C_{\mathbf K,\alpha,\beta}$ annihilates a photon in the indicated mode. Restricting to zero radial index and to the near-waist region $|z|\ll z_R$, the spatial mode is taken as~\cite{LOU1,Allen1992OAM}
\begin{equation}
U_{kl}(\mathbf r)
={}
\frac{1}{\sqrt{\pi |l|!}}
\left(
\frac{\sqrt{2}}{w_0}
\right)^{|l|+1}
\rho^{|l|}
\exp\left[
-\frac{\rho^2}{w_0^2}
+
\frac{ikz\rho^2}{2z_R^2}
+
il\phi
-
i(|l|+1)\frac{z}{z_R}
\right],
\label{eq:lg_mode_original}
\end{equation}
where $\mathbf r=(\rho,\phi,z)$ is expressed in cylindrical coordinates, the Rayleigh range is
\begin{equation}
z_R=\frac{1}{2}kw_0^2,
\end{equation}
and $l=0,\pm1,\pm2,\ldots$.

In second quantization, the electron--light interaction Hamiltonian is
\begin{equation}
\mathcal H_{\mathrm{int}}
=
\int d^3r\,
\hat\Psi^{\dagger}(\mathbf r)
H_{\mathrm{int}}
\hat\Psi(\mathbf r),
\label{eq:second_quantized_interaction}
\end{equation}
where the electron field operator is expanded as
\begin{equation}
\hat\Psi(\mathbf r)
=
\frac{1}{\sqrt V}
\sum_{n,\mathbf p}
\phi_{n\mathbf p}(\mathbf r)
\hat e_{n\mathbf p},
\label{eq:field_operator}
\end{equation}
with
\begin{equation}
\phi_{n\mathbf p}(\mathbf r)
=
\exp(i\mathbf p\cdot\mathbf r)
 u_{n\mathbf p}(\mathbf r).
\label{eq:bloch_function}
\end{equation}
Here $u_{n\mathbf p}(\mathbf r)$ is the periodic part of the Bloch function. Retaining only the $\Gamma_7^+$ valence band and $\Gamma_6^+$ conduction band, with $n=v,c$, gives
\begin{equation}
\mathcal H_{\mathrm{int}}
=
-\frac{e}{m_0c}
A_0
\left(
\alpha\mathbf e_x+\beta\mathbf e_y
\right)
\cdot
\mathbf P^{\mathrm{LG}}_{cv}
\hat C^{\dagger}_{\mathbf k}
\hat e_{\mathbf p}
\hat h_{\mathbf q}
\delta(\mathbf p-\mathbf k,\mathbf q)
+
\mathrm{H.c.}
\label{eq:interaction_band_basis}
\end{equation}
The interband matrix element is
\begin{equation}
\mathbf P^{\mathrm{LG}}_{cv}
={}
\frac{1}{V}
\int d^3r\,
e^{-i\mathbf q\cdot\mathbf r}
 u_{v}^{*}(\mathbf r)
U_{kl}(\mathbf r)
e^{-ikz}
(-i\boldsymbol\nabla)
e^{i\mathbf p\cdot\mathbf r}
 u_c(\mathbf r).
\label{eq:pcv_general}
\end{equation}
After expanding the Bloch functions around zero wave vector, the matrix element takes the form
\begin{equation}
\mathbf P^{\mathrm{LG}}_{cv}
={}
\frac{1}{\Omega}
\int_{\Omega_0}
 d^3r\,
 u_{v0}^{*}
U_{kl}
(-i\boldsymbol\nabla)
 u_{c0}
+
\frac{\mathbf p}{\Omega_0}
\int_{\Omega_0}
 d^3r\,
 u_{v0}^{*}
U_{kl}
 u_{c0}
+
M_{\alpha\beta}^{cv}p_{\alpha},
\label{eq:pcv_expansion}
\end{equation}
where $M_{\alpha\beta}^{cv}$ is the two-photon transition tensor,
\begin{equation}
M_{\alpha\beta}^{cv}
={}
\sum_i
\frac{\hbar p_i}{m_0}
\Bigg[
\frac{
\langle u_{v0}U_{kl}|p_{\alpha}|u_{i0}\rangle
\langle u_{i0}U_k|p_{\beta}|u_{c0}\rangle
}{E_c-E_i}
+
\frac{
\langle u_{v0}U_k|p_{\alpha}|u_{i0}\rangle
\langle u_{i0}U_{kl}|p_{\beta}|u_{c0}\rangle
}{E_v-E_i}
\Bigg].
\label{eq:two_photon_tensor}
\end{equation}

For ordinary light, the first term in Eq.~\eqref{eq:pcv_expansion} vanishes because the conduction- and valence-band functions have the same parity, and the second term vanishes by orthogonality. The transition is therefore forbidden. In an LG field, however, the additional spatial factor $U_{kl}$ changes the symmetry of the integrand. Because the LG order $l$ can be varied, the first two terms in Eq.~\eqref{eq:pcv_expansion} can become nonzero for particular OAM channels.
\section{Paraexciton State and LG Selection Channels}
\label{sec:paraexciton}

Within the effective-mass approximation, the exciton wavefunction is written as
\begin{equation}
\Psi^{\mathrm{Ex}}
=
\Phi_{\nu}(\mathbf r_e-\mathbf r_h)
 u_v(\mathbf r_h)
 u_c(\mathbf r_e),
\label{eq:exciton_wavefunction_real_space}
\end{equation}
where $\Phi_{\nu}(\mathbf r_e-\mathbf r_h)$ is the envelope function of the relative electron--hole motion, and $u_v$ and $u_c$ are the valence- and conduction-band Bloch functions.

The yellow exciton series of cuprous oxide is formed from electrons and holes in the lowest conduction band $\Gamma_6^+$ and highest valence band $\Gamma_7^+$ of the $O_h$ symmetry group. The minimum of $\Gamma_6^+$ and the maximum of $\Gamma_7^+$ occur at the same point in momentum space, with an energy gap of approximately $2.17\,\mathrm{eV}$. Because $\Gamma_6^+$ and $\Gamma_7^+$ have the same parity, the interband electric-dipole matrix element vanishes.

The product representation is
\begin{equation}
\Gamma_6^+\otimes\Gamma_7^+
=
\Gamma_2^+\oplus\Gamma_5^+.
\end{equation}
Accordingly, the four $1S$ yellow-exciton states split into a triplet orthoexciton of $\Gamma_5^+$ symmetry, with $J=1$ and $J_z=0,\pm1$, and a singlet paraexciton of $\Gamma_2^+$ symmetry, with $J=0$ and $J_z=0$. Although these states are conventionally called orthoexciton and paraexciton states, they are total-angular-momentum states and are not pure spin states.

The $\Gamma_5^+$ orthoexciton has basis functions transforming as $yz$, $xz$, and $xy$~\cite{KOS1,Dresselhaus2008}, allowing electric-quadrupole transitions. The $\Gamma_2^+$ paraexciton has a basis function proportional to
\begin{equation}
F_{\Gamma_2}(x,y,z)
=
(x^2-y^2)(y^2-z^2)(z^2-x^2),
\label{eq:gamma2_polynomial}
\end{equation}
and is therefore highly forbidden in the usual optical channels. The paraexciton state is written as
\begin{equation}
|J=0,J_z=0\rangle
={}
\frac{1}{\sqrt{2}}
F_{\Gamma_2}(x,y,z)
\left(
|\downarrow_e,\uparrow_h\rangle
-
|\uparrow_e,\downarrow_h\rangle
\right).
\label{eq:paraexciton_state}
\end{equation}
The degree-six orbital factor in Eq.~\eqref{eq:paraexciton_state} motivates the coupling to high-order LG modes.

Substituting the paraexciton state into Eq.~\eqref{eq:pcv_expansion} gives two nonzero symmetry channels. First, the gradient term is nonzero for $l=5$,
\begin{equation}
\mathbf P_{cv}^{\mathrm{LG},l=5}
=
\frac{1}{\Omega_0}
\int_{\Omega_0}
 d^3r\,
 u_{v0}^{*}
U_{k5}
(-i\boldsymbol\nabla)
 u_{c0}.
\label{eq:pcv_l5_channel}
\end{equation}
Second, the overlap term is nonzero for $l=6$,
\begin{equation}
\mathbf P_{cv}^{\mathrm{LG},l=6}
=
\frac{\mathbf p}{\Omega_0}
\int_{\Omega_0}
 d^3r\,
 u_{v0}^{*}
U_{k6}
 u_{c0}.
\label{eq:pcv_l6_channel}
\end{equation}
Thus, the paraexciton becomes optically accessible for LG beams of order $l=5$ or $l=6$.

For $l=6$, the matrix element is written as
\begin{equation}
\mathbf P_{cv}^{l=6}
={}
\frac{\mathbf p}{\Omega_0\sqrt{2}}
\left\langle
U_{k6}
F_{\Gamma_2}(x,y,z)
\right\rangle
\left(
|\downarrow_e,\uparrow_h\rangle
-
|\uparrow_e,\downarrow_h\rangle
\right).
\label{eq:pcv_l6_projection}
\end{equation}
Projecting the cylindrical coordinates of the LG beam onto the spherical coordinates used for the exciton state gives the original result
\begin{equation}
\mathbf P_{cv}^{\mathrm{LG},l=6}
={}
\frac{\mathbf p}{\Omega_0 k\omega_0^2}
\frac{3\sqrt{3}}{6!\pi}
 r^9\exp(-r^{2})
\left(
|\downarrow_e,\uparrow_h\rangle
-
|\uparrow_e,\downarrow_h\rangle
\right),
\label{eq:pcv_l6_original_result}
\end{equation}
where $r$ ranges over the unit-cell radius.

Similarly, for $l=5$,
\begin{equation}
\mathbf P_{cv}^{\mathrm{LG},l=5}
={}
\frac{1}{\Omega_0}
\left\langle
U_{k5}
(-i\boldsymbol\nabla)
F_{\Gamma_2}(x,y,z)
\right\rangle
\left(
|\downarrow_e,\uparrow_h\rangle
-
|\uparrow_e,\downarrow_h\rangle
\right),
\label{eq:pcv_l5_projection}
\end{equation}
with the original projected result
\begin{equation}
\mathbf P_{cv}^{\mathrm{LG},l=5}
={}
\frac{1}{\Omega_0 k\omega_0^2}
\frac{3\sqrt{2}}{5!\pi}
 r^7\exp(-r^{2})
\left(
|\downarrow_e,\uparrow_h\rangle
-
|\uparrow_e,\downarrow_h\rangle
\right).
\label{eq:pcv_l5_original_result}
\end{equation}
Equations~\eqref{eq:pcv_l6_original_result} and~\eqref{eq:pcv_l5_original_result} give the projected matrix elements for the two symmetry-allowed channels.
\section{Transition Matrix Element}
\label{sec:matrix_element}

In the second-quantized representation, the exciton state is written as
\begin{equation}
|\Psi_{\nu}^{\mathrm{Ex}}(\mathbf K)\rangle
=
\sum_{\mathbf p}
\Phi_{\nu}(\mathbf p)
\hat e^{\dagger}(\beta_x\mathbf K+\mathbf p)
\hat h^{\dagger}(\alpha_x\mathbf K-\mathbf p)
|0\rangle,
\label{eq:exciton_state_momentum}
\end{equation}
where
\begin{equation}
\alpha_x=\frac{m_h}{m_e+m_h},
\qquad
\beta_x=1-\alpha_x,
\end{equation}
and $m_e$ and $m_h$ are the electron and hole effective masses, respectively. The function $\Phi_{\nu}(\mathbf p)$ is the Fourier transform of the electron--hole envelope function. In real space,
\begin{equation}
\Phi_{\nu}(\mathbf r)
=
R_{nl}(r)
Y_{lm}(\hat{\mathbf r}).
\end{equation}
For the $1S$ exciton, the original dimensionless momentum-space envelope is
\begin{equation}
\Phi_{1S}(\mathbf p)
=
\frac{8\sqrt{\pi}}{(1+p^2)^2}.
\label{eq:phi_1s_momentum}
\end{equation}
Introducing the electron and hole wave vectors
\begin{equation}
\mathbf p_e=\beta_x\mathbf K+\mathbf p,
\qquad
\mathbf p_h=\alpha_x\mathbf K-\mathbf p,
\end{equation}
Eq.~\eqref{eq:exciton_state_momentum} becomes
\begin{equation}
|\Psi_{1S}^{\mathrm{Ex}}(\mathbf K)\rangle
={}
8\sqrt{\pi}
\sum_{\mathbf p_e,\mathbf p_h}
\frac{
\delta(\mathbf K-\mathbf p_e-\mathbf p_h)
}{
\left[1+(\alpha_x\mathbf K-\mathbf p_h)^2\right]^2
}
\times
\hat e^{\dagger}(\mathbf p_e)
\hat h^{\dagger}(\mathbf p_h)
|0\rangle.
\label{eq:exciton_state_1s}
\end{equation}
The transition matrix element for creation of the paraexciton by the LG field is therefore
\begin{equation}
\langle \mathrm{Ex}_{\mathrm{para}}|\mathcal H_{\mathrm{int}}|0\rangle
={}
-\frac{e}{m_0c}
\sum_{\mathbf p_h}
\frac{1}{
\left[1+(\alpha_x\mathbf K-\mathbf p_h)^2\right]^2
}
\times
\left(
\alpha\mathbf e_x+\beta\mathbf e_y
\right)
\cdot
\mathbf P_{cv}^{\mathrm{LG}}(\mathbf K-\mathbf p_h).
\label{eq:paraexciton_transition_matrix_element}
\end{equation}
This expression completes the original symmetry-based construction. It establishes that the paraexciton matrix element can be nonzero for the $l=5$ and $l=6$ channels. The remaining question is quantitative: whether a spatially realistic LG field produces appreciable overlap with the compact $1S$ exciton. That question motivates the finite-waist analysis developed next.

\section{Spatial Localization and Optimal OAM Coupling}
\label{sec:spatial_localization}

The symmetry analysis establishes that the $1S$ paraexciton can be reached through a direct $l=6$ channel and a gradient-assisted $l=5$ channel. The remaining question is whether these symmetry-allowed matrix elements remain appreciable once the optical and excitonic length scales are included. We address that question in three steps. We first evaluate the two channels for finite-waist Laguerre--Gaussian beams, then replace the LG profile by a broader class of localized OAM fields, and finally determine how the optimal localization scale changes with the polynomial degree of the target cubic harmonic.

\subsection{Finite-waist Laguerre--Gaussian beams}
\label{sec:finite_lg}

For a conventional LG beam, the transverse scale is set by the beam waist $w_0$. We evaluate the direct $l=6$ and gradient-assisted $l=5$ overlap integrals as functions of $w_0/a_B$, where $a_B$ is the exciton Bohr radius.

\begin{figure}[t]
    \centering
    \includegraphics[width=\linewidth]
    {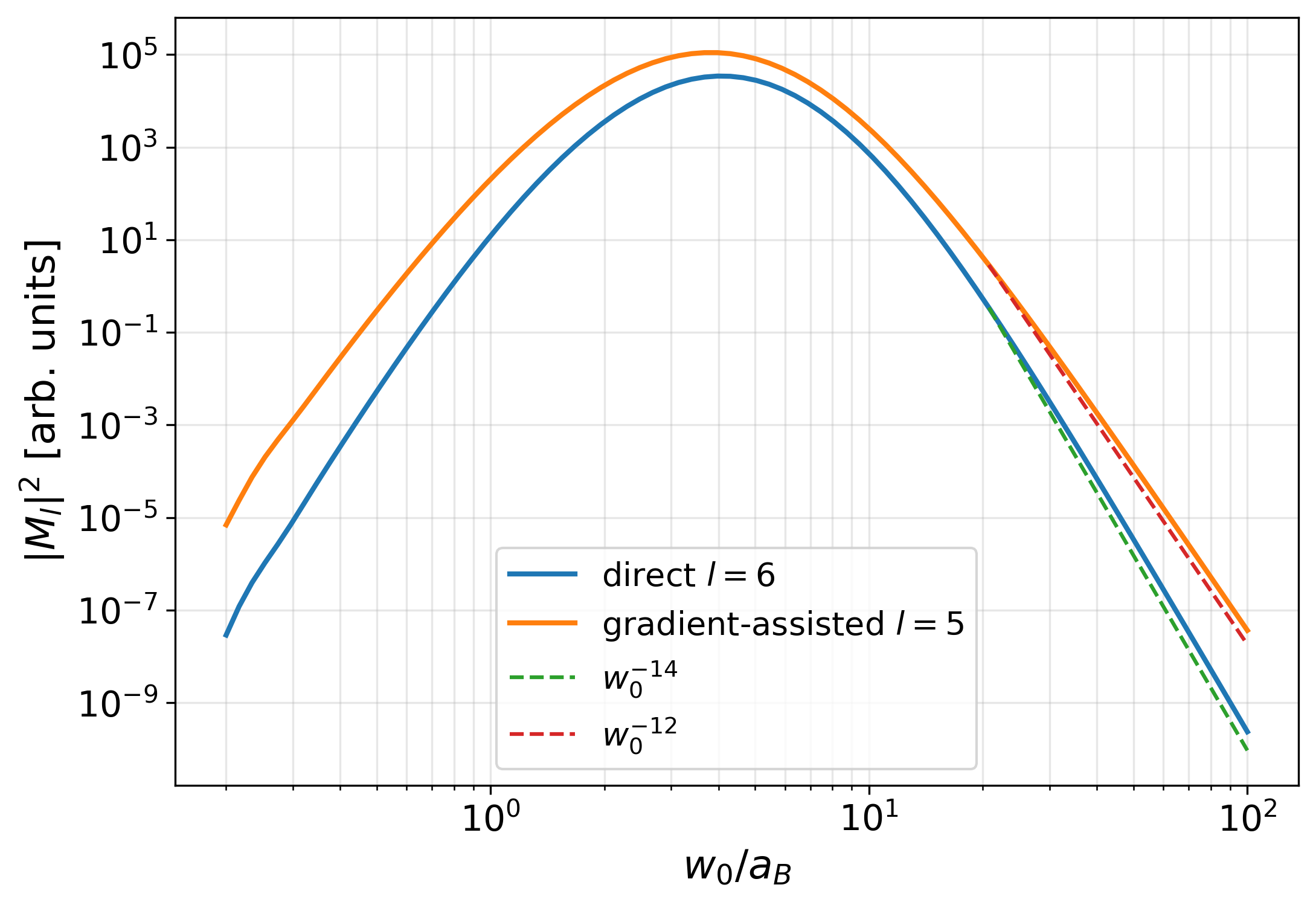}
    \caption{Squared coupling strengths $|M_{l}|^2$ for the direct $l=6$ and gradient-assisted $l=5$ channels as a function of the normalized LG beam waist $w_0/a_B$.
    The coupling decreases rapidly once the transverse scale of the beam
    becomes large compared with the exciton Bohr radius.}
    \label{fig:finite_waist_coupling}
\end{figure}

Figure~\ref{fig:finite_waist_coupling} shows that both channels are appreciable only when the field varies on a scale comparable to that of the exciton. At large waist, the squared coupling strengths scale as
\begin{equation}
|M_{l=6}|^2
\propto
\left(\frac{a_B}{w_0}\right)^{14}
\end{equation}
and
\begin{equation}
|M_{l=5}|^2
\propto
\left(\frac{a_B}{w_0}\right)^{12}.
\end{equation}
These powers follow from the near-axis form of a normalized LG field,
\begin{equation}
U_l(\rho,\phi)
\propto
\frac{1}{w_0}
\left(\frac{\rho}{w_0}\right)^{|l|}
e^{il\phi}.
\end{equation}
The factor $\rho^{|l|}$ produces the vortex core, while the additional factor $1/w_0$ follows from normalization at fixed optical power \cite{Allen1992OAM}. When $w_0\gg a_B$, the exciton samples only the weak near-axis portion of the beam. Thus, satisfying the angular selection rule is not sufficient: the required angular structure must also be delivered on the internal exciton length scale.

\subsection{Localized OAM fields and the intensity-ring radius}
\label{sec:localized_oam}

To separate the azimuthal winding from the radial confinement, we consider the generalized field
\begin{equation}
U_{kl}(\rho,\phi,z)
=
\mathcal N_l(\xi)
\left(\frac{\rho}{\xi}\right)^{|l|}
f(\rho,z;\xi)
e^{il\phi},
\label{eq:general_localized_oam}
\end{equation}
where $\xi$ sets the characteristic localization length, $\mathcal N_l(\xi)\propto\xi^{-3/2}$ sets the overall field normalization, and $f(\rho,z;\xi)$ specifies the radial and longitudinal envelope. The index $k$ is retained for consistency with the preceding sections, although the present calculation depends only on the transverse structure.

We use exponential, Gaussian, and super-Gaussian envelopes,
\begin{equation}
f(\rho;\xi)
=
\exp\left[-\left(\frac{\rho}{\xi}\right)^p\right],
\label{eq:localized_profile}
\end{equation}
with $p=1$, $p=2$, and $p=4$, respectively. These are model profiles rather than specific experimental sources. Their purpose is to test which conclusions depend on the detailed radial shape and which follow more generally from localized OAM.

The parameter $\xi$ does not identify the same physical feature for all three profiles. At fixed $\xi$, the radius of maximum intensity depends on both $p$ and $l$, so coupling curves plotted against $\xi/a_B$ peak at different apparent localization lengths. A more direct geometric variable is the intensity-ring radius $\rho_{\rm peak}$. For a field amplitude proportional to
\begin{equation}
\left(\frac{\rho}{\xi}\right)^{|l|}
\exp\left[-\left(\frac{\rho}{\xi}\right)^p\right],
\end{equation}
the intensity maximum occurs at
\begin{equation}
\rho_{\rm peak}
=
\xi
\left(\frac{|l|}{p}\right)^{1/p}.
\label{eq:rho_peak_general}
\end{equation}

We therefore reparameterize the coupling curves by replacing $\xi/a_B$ with $\rho_{\rm peak}/a_B$. The calculated coupling values are unchanged; only the horizontal coordinate is transformed.

\begin{figure}[t]
\centering
\includegraphics[width=\linewidth]{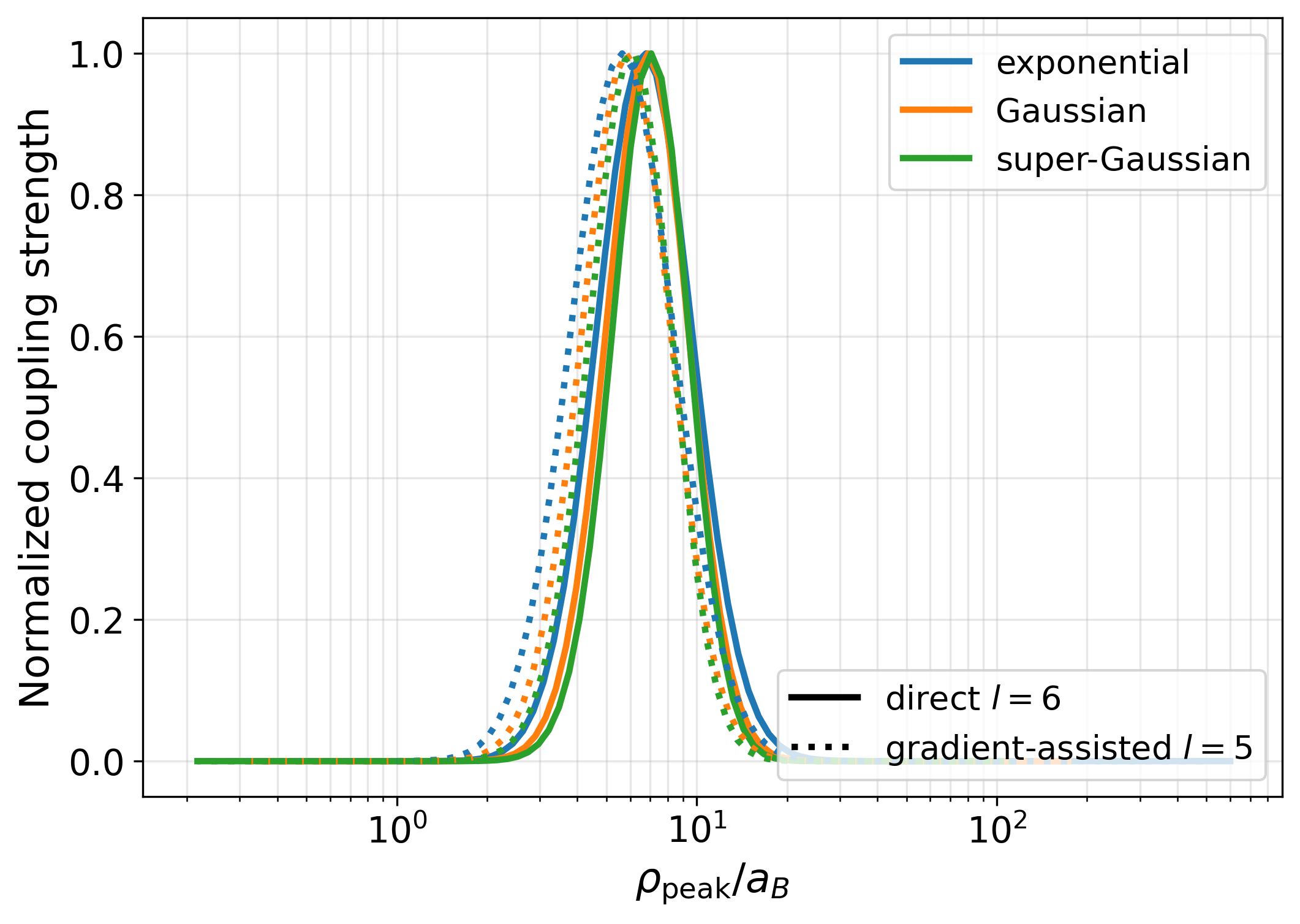}
\caption{Normalized coupling strengths as a function of the physical OAM intensity-ring radius $\rho_{\rm peak}/a_B$ for the exponential, Gaussian, and super-Gaussian localized OAM profiles. Solid curves denote the direct $l=6$ channel, while dotted curves denote the gradient-assisted $l=5$ channel. The two channels peak near $\sim 7a_B$ and $\sim 6a_B$, respectively, with substantially reduced dependence on the detailed radial profile when expressed in terms of $\rho_{\rm peak}$.}
\label{fig:rho_peak_collapse}
\end{figure}

Figure~\ref{fig:rho_peak_collapse} shows that both channels align much more closely when expressed in terms of $\rho_{\rm peak}/a_B$. Most of the apparent profile dependence in scans versus $\xi/a_B$ therefore comes from comparing fields whose bright rings lie at different radii. The remaining differences are smaller corrections associated with the detailed radial envelope. The intensity-ring radius is consequently the natural variable for comparing the optical and excitonic length scales.

\subsection{Degree scaling and profile robustness}
\label{sec:degree_scaling}

We next ask how the optimal ring radius changes with the polynomial degree of the target cubic harmonic. We begin with a Gaussian localized OAM profile, $p=2$, and homogeneous cubic-harmonic polynomials of degree $n=1,\ldots,6$. Here homogeneous means that every term in a given polynomial has the same total power of the Cartesian coordinates. For each representative, we choose $l=n$, scan $\xi/a_B$ over the range $0.2$--$100$, and evaluate $|M_{n,n}(\xi)|^2$. The Gaussian localization parameter is converted to
\begin{equation}
\rho_{\rm peak}
=
\xi
\sqrt{\frac{n}{2}},
\label{eq:rho_peak_degree_gaussian}
\end{equation}
and each curve is normalized by its own maximum. The optimal radius $\rho_{\rm opt}$ is the sampled value of $\rho_{\rm peak}$ at which the coupling is largest.

A simple radial estimate follows from
\begin{equation}
W_n(r)
=
r^n e^{-r/a_B},
\label{eq:degree_radial_factor}
\end{equation}
which combines the near-origin suppression of a degree-$n$ polynomial with the decay of the hydrogenic $1S$ envelope used in the present overlap model. Detailed treatments of the low-lying Cu$_2$O excitons include central-cell, band-structure, and cubic-symmetry corrections to this simple hydrogenic description \cite{heckotter2025}. Its maximum occurs at $r = n a_B$, suggesting that the optimal OAM ring should move outward approximately linearly with $n$.

\begin{figure}[t]
\centering
\includegraphics[width=\linewidth]{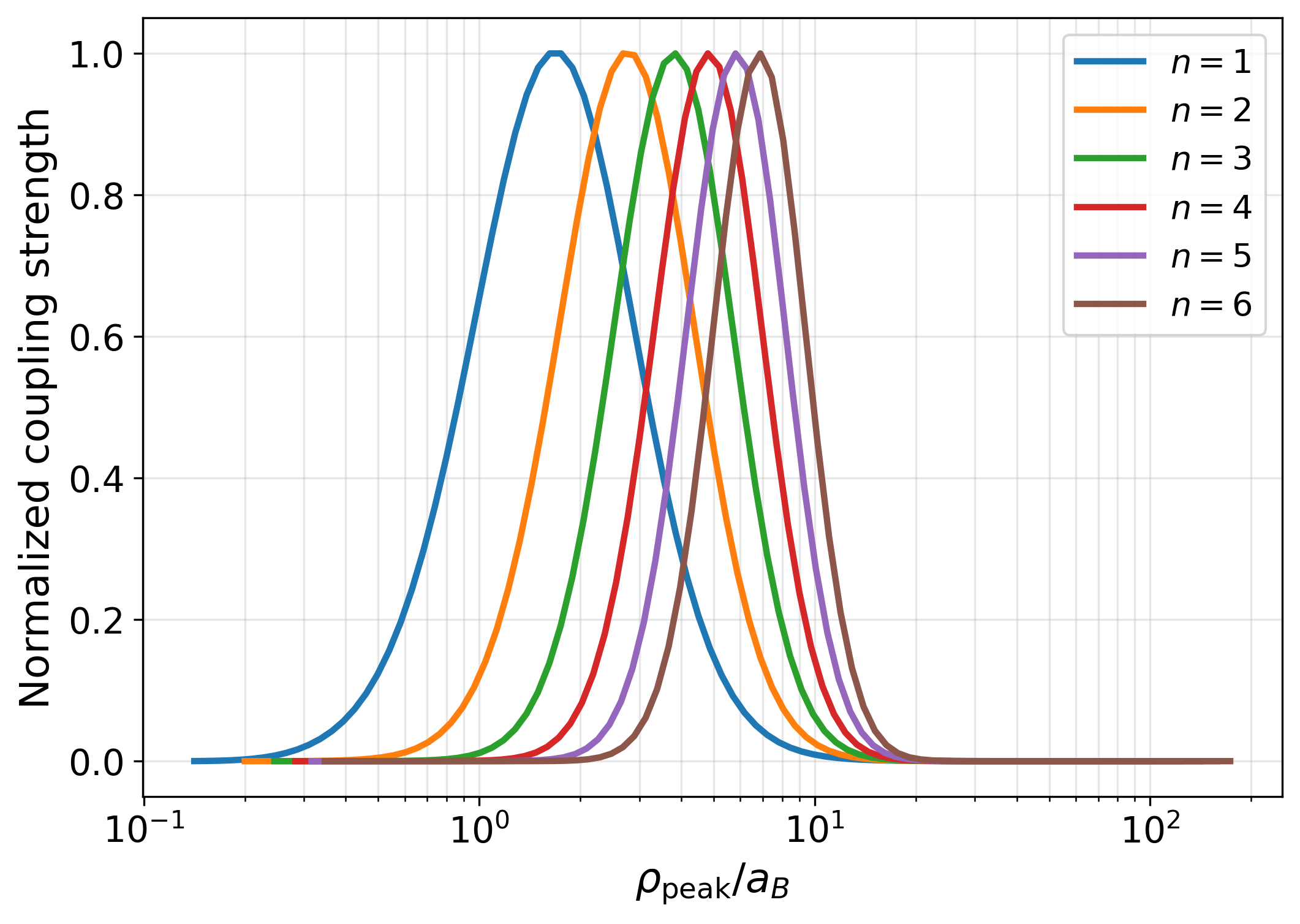}
\caption{Normalized coupling strengths for cubic-harmonic representatives of degree $n=1,\ldots,6$, calculated using a Gaussian localized OAM profile with $l=n$. The optimal radius shifts systematically outward as the polynomial degree increases.}
\label{fig:degree_scaling_curves}
\end{figure}

\begin{figure}[t]
\centering
\includegraphics[width=\linewidth]{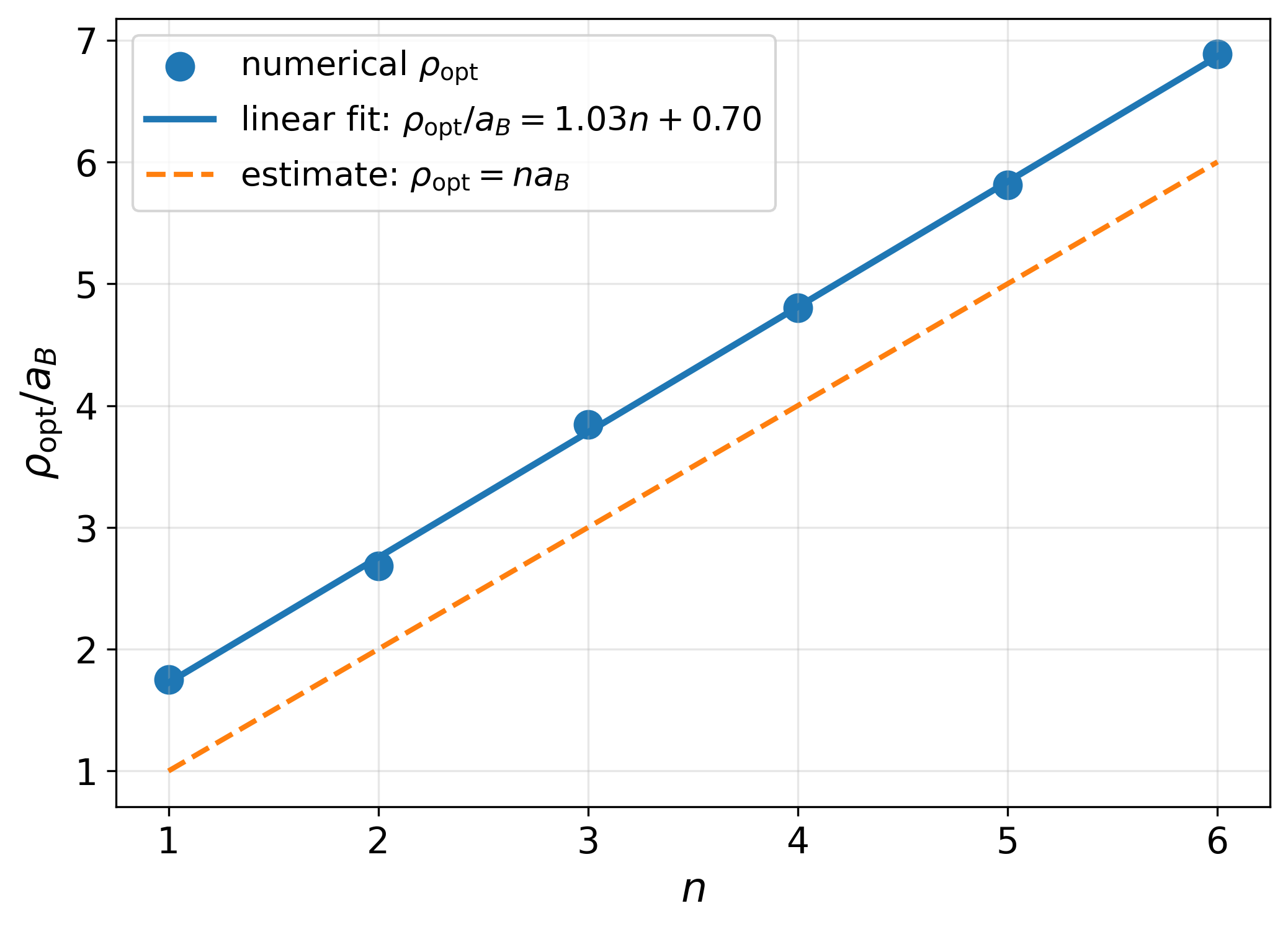}
\caption{Optimal intensity-ring radius $\rho_{\rm opt}/a_B$ as a function of the polynomial degree $n$ for cubic-harmonic representatives and a Gaussian localized OAM profile.}
\label{fig:degree_scaling_fit}
\end{figure}

The numerical curves in Fig.~\ref{fig:degree_scaling_curves} follow this estimate. Increasing $n$ suppresses the target amplitude more strongly near the origin and shifts the dominant overlap toward larger radii. The extracted optimal radii are shown in Fig.~\ref{fig:degree_scaling_fit}.

A linear fit gives
\begin{equation}
\frac{\rho_{\rm opt}}{a_B}
=
1.0284\,n+0.6964,
\label{eq:gaussian_degree_fit}
\end{equation}
so the optimum increases by approximately one Bohr radius for each unit increase in polynomial degree. The degree-six representative in Eq.~\eqref{eq:gamma2_polynomial} corresponds to the $\Gamma_2^+$ paraexciton symmetry factor up to an overall sign \cite{Dresselhaus2008}.

\begin{figure}[t]
\centering
\includegraphics[width=\linewidth]{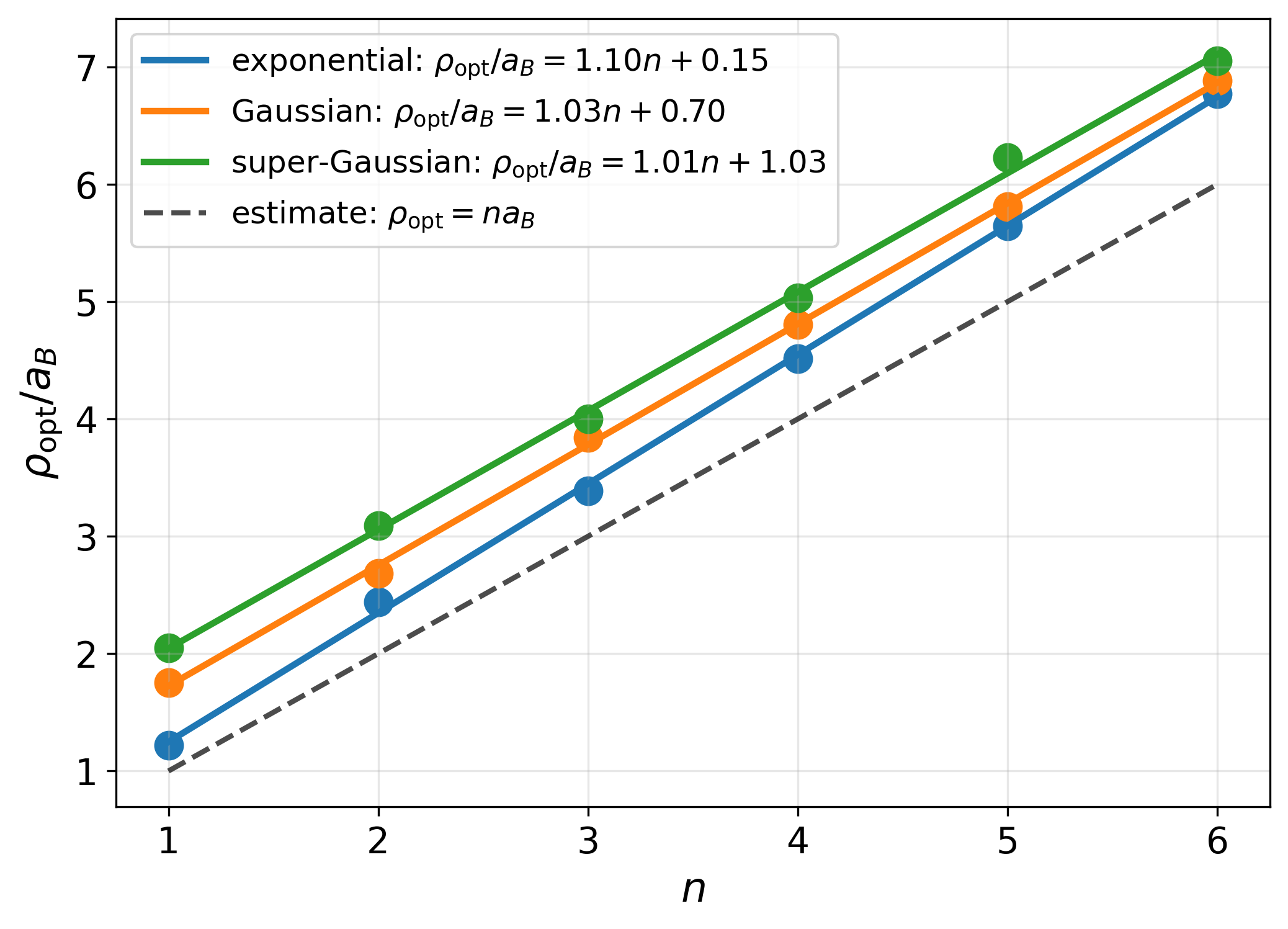}
\caption{Optimal intensity-ring radius $\rho_{\rm opt}/a_B$ as a function of polynomial degree for exponential, Gaussian, and super-Gaussian localized OAM profiles. The fitted slopes remain close to unity, while the main profile dependence appears in the intercept.}
\label{fig:profile_scaling_fit}
\end{figure}

To test the robustness of the scaling, we repeat the calculation for exponential, Gaussian, and super-Gaussian profiles while retaining the same cubic-harmonic representatives and the choice $l=n$, shown in Fig.~\ref{fig:profile_scaling_fit}. For each profile and degree, the extracted optimal radii are fitted to
\begin{equation}
\frac{\rho_{\rm opt}}{a_B}
=
\alpha n+\beta,
\label{eq:profile_degree_fit}
\end{equation}
where $\alpha$ gives the leading degree dependence and $\beta$ gives the profile-dependent offset.

\begin{figure}[t]
\centering
\includegraphics[width=\linewidth]{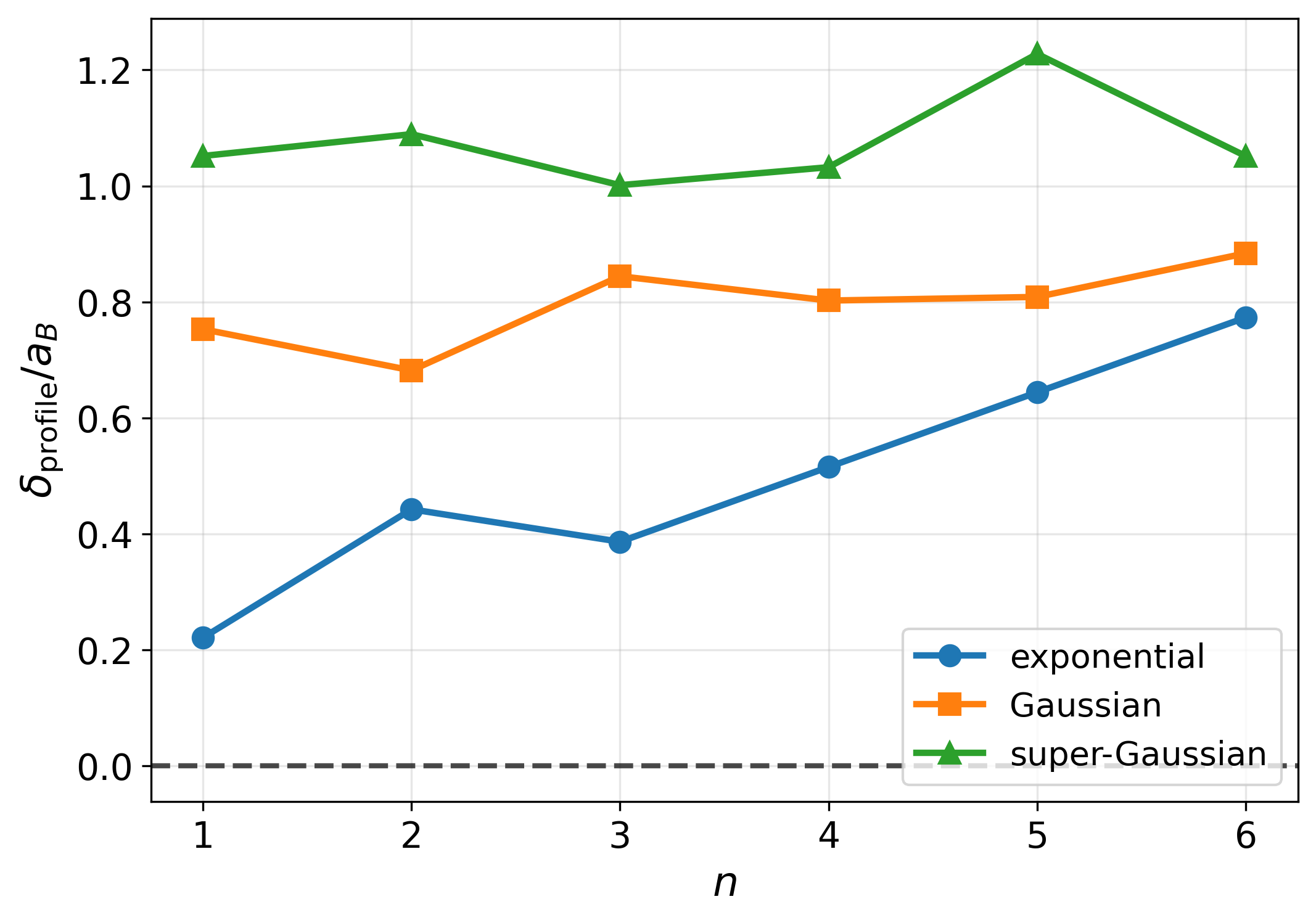}
\caption{Profile-dependent correction $\delta_{\rm profile}=\rho_{\rm opt}-n a_B$, shown in units of the exciton Bohr radius $a_B$, for the exponential, Gaussian, and super-Gaussian localized OAM profiles.}
\label{fig:profile_delta}
\end{figure}

The slopes remain close to unity for all three profiles. The detailed radial envelope therefore shifts the precise location of the optimum without changing the leading linear dependence on $n$. We write the deviation from $\rho_{\rm opt}=n a_B$ as $\delta_{\rm profile} = \rho_{\rm opt} - n a_B$ and plot it in Fig.~\ref{fig:profile_delta} for each of the three test profiles.

The calculations support the approximate scaling
\begin{equation}
    \rho_{\rm opt} \approx n a_B + \delta_{\rm profile},
\label{eq:optimal_rho}
\end{equation}
with $\delta_{\rm profile}$ remaining of order $a_B$ over the degrees considered. For the degree-six $\Gamma_2^+$ paraexciton factor, all three profiles place the optimal intensity ring near $7a_B$. The cubic symmetry determines which OAM channels are allowed, while the polynomial degree sets the leading radial scale at which the coupling is strongest.

\section{Discussion and Outlook}
\label{sec:discussion}

The results point to two separate requirements for direct OAM-assisted excitation of the paraexciton. The first is angular. The optical field must contain the spatial phase structure needed to match the $\Gamma_2^+$ exciton symmetry, which produces the allowed $l=5$ and $l=6$ channels derived above. The second is radial. Satisfying the angular selection rule does not by itself guarantee an appreciable transition. The field must also vary over the spatial extent of the exciton and place substantial optical weight in the region where the exciton overlap is largest.

This distinction explains the strong suppression found for conventional far-field Laguerre--Gaussian beams. Such a beam can carry the correct winding while still varying too slowly across the internal exciton coordinate. In that limit, the optical phase changes little over the exciton, and the higher-order spatial structure required by the paraexciton remains weak. The problem is therefore not the absence of the correct angular momentum, but the mismatch between the optical and excitonic length scales.

Localized OAM fields provide a route to overcoming this mismatch by allowing the vortex structure to be confined independently of the free-space wavelength. For the degree-six $\Gamma_2^+$ paraexciton factor, the numerical calculations place the optimal intensity ring near $\rho_{\rm opt} \sim 6a_B\text{--}7a_B$. Reported values of the yellow $1S$ exciton Bohr radius in Cu$_2$O lie near $0.7$--$0.8~\mathrm{nm}$ \cite{Tayagaki2005}. The corresponding physical ring radius is therefore approximately $\rho_{\rm opt} \sim 4\text{--}5.5~\mathrm{nm}$. This estimate makes clear that the required localization is well beyond the diffraction limit \cite{hell2007}. LG beams do not circumvent the diffraction limit; their phase-matching conditions are identical to those of plane waves \cite{yao2011}. Standard focused beams and conventional dielectric cavity modes are generally limited to diffraction-scale confinement \cite{Saldutti2021}, and therefore remain too broad to reproduce the few-nanometer internal spatial structure required here.

Plasmonic near fields provide a more plausible route to this regime. Surface-plasmon vortices and metasurface-generated OAM fields demonstrate that optical angular momentum can be created and manipulated below the diffraction limit \cite{Chen2015NearFieldOAM}. Plasmonic nanogaps and picocavities provide still stronger localization, with field confinement extending to nanometer and, in some cases, subnanometer scales \cite{Zhang2017Subnanometre}. These results do not by themselves establish that a suitable $l=5$ or $l=6$ vortex can be produced at the required radius. They do show, however, that the relevant spatial scale is not outside the range of present nanophotonic structures.

The required source is more specific than a generic optical hot spot. A strongly confined field with no controlled azimuthal phase does not satisfy the paraexciton symmetry condition. Conversely, a well-defined optical vortex with a ring radius of tens or hundreds of nanometers remains too large. The useful source must combine both properties: the correct winding and an intensity ring with a radius of only a few nanometers. Candidate structures therefore include rotationally structured plasmonic nanogaps, vortex-supporting nanoparticle-on-mirror cavities, and related near-field geometries in which the phase and radial confinement can be designed together.

The present calculation establishes the symmetry channels and the geometric scale required for strong overlap. It does not yet provide an absolute excitation rate. The localized profiles used here specify the spatial form of the field, but their amplitudes are normalized independently of a particular source. A quantitative comparison with experiment will require the full vector electromagnetic field of a realizable nanostructure, including its normalization, frequency dependence, and orientation relative to the Cu$_2$O crystal axes. The field inside the semiconductor must also be obtained rather than inferred directly from the field in vacuum. A more quantitative treatment should also replace the hydrogenic $1S$ envelope by a Cu$_2$O exciton wavefunction including the central-cell and band-structure corrections that are important for compact low-lying excitons \cite{rommel2021exchange}.

Loss and linewidth will be equally important. Metallic localization increases the field strength, but it also introduces absorption and broad plasmonic resonances \cite{Boriskina2017PlasmonicLosses}. A useful structure must produce sufficient field at the paraexciton energy without broadening or heating the system enough to remove the advantage of direct excitation. The calculation must therefore be extended from the dimensionless overlap considered here to a transition rate containing the local field amplitude, the electromagnetic density of states, and the paraexciton linewidth. That rate can then be compared with competing processes such as phonon-assisted excitation, nonradiative loss, and indirect population through the orthoexciton manifold \cite{Schoene2017PhononAssisted,Kubouchi2005OrthoPara}.

The degree scaling found in Sec.~\ref{sec:degree_scaling} also suggests a broader interpretation. For an excitonic target whose leading spatial symmetry is represented by a polynomial of degree $n$, the near-origin amplitude is suppressed by the corresponding power of the internal coordinate. The dominant radial overlap then moves outward with increasing $n$, producing the approximate rule in Eq.~\eqref{eq:optimal_rho}, where the profile-dependent correction remains of order $a_B$ for the fields considered here. The cubic symmetry determines which optical winding is allowed, while the polynomial degree determines the leading radial scale at which that allowed coupling becomes strongest. These are related requirements, but they are not interchangeable.

\section{Conclusion}
\label{sec:conclusion}

We have shown that the nominally forbidden $1S$ paraexciton in Cu$_2$O can acquire nonzero optical matrix elements when the incident field carries the angular structure required by the $\Gamma_2^+$ exciton symmetry. The symmetry analysis identifies the $l=5$ and $l=6$ OAM channels as the leading allowed contributions. These channels follow from the symmetry construction developed above and do not depend on the detailed radial form of the optical field.

The finite-waist calculation shows, however, that an allowed transition need not be an appreciable one. For a conventional Laguerre--Gaussian beam with a waist much larger than the exciton, the field varies only weakly over the internal exciton coordinate. The resulting matrix element is therefore strongly suppressed even when the OAM winding satisfies the symmetry requirement. This separates the problem into two conditions: the field must have the correct angular content, and it must vary on the spatial scale of the exciton.

Localized OAM fields provide a way to satisfy both conditions. Expressing the coupling in terms of the physical intensity-ring radius $\rho_{\rm peak}$ removes the profile-dependent ambiguity associated with the localization parameter $\xi$ and gives a direct comparison between the optical and excitonic length scales. The numerical results show that the optimal ring radius is governed primarily by the polynomial degree $n$ of the target cubic harmonic. Over the profiles considered here, the optimum follows the approximate scaling of Eq.~\ref{eq:optimal_rho}, where the profile-dependent correction remains of order $a_B$. For the degree-six $\Gamma_2^+$ paraexciton factor, the optimal ring lies near $6a_B$--$7a_B$.

The main result is therefore not simply that OAM can relax the paraexciton selection rule. The field must also be localized so that its angular structure is sampled over the radial region in which the exciton overlap is largest. The cubic symmetry determines which OAM channels are allowed, while the polynomial degree sets the leading radial scale for strong coupling.

The present calculation establishes the symmetry channels, the finite-waist suppression, and the localized-field design rule within the overlap model. A quantitative prediction for experiment will require the full vector field of a realizable nanophotonic source, together with absolute field normalization, material response, loss, and linewidth. Those ingredients will determine whether the allowed matrix elements found here produce an experimentally useful paraexciton excitation rate.
\begin{acknowledgments}
N.Q.H. is thankful for financial support from grant RCG23-007 (WVURC-MURC 23-049).
\end{acknowledgments}

\bibliography{references_verified}

\end{document}